\documentclass{article}
\usepackage{url}
\usepackage{float}
\usepackage{comment}
\usepackage{setspace}
\usepackage{blindtext}
\usepackage{graphicx}
\usepackage{amsmath}
\usepackage{amsfonts}
\usepackage{lipsum}
\usepackage{wrapfig}
\usepackage{multirow}

\begin{document}

\title{\large\bf An Efficient Detection of Malware by Naive Bayes Classifier Using GPGPU\footnote{{\bf Reference:} Springer, Advances in Computer Communication and Computational Sciences, Vol. 924, pp. 255-262, 2019}}
\author{\normalsize Sanjay K. Sahay and Mayank Chaudhari \\ {\normalsize BITS, Pilani, Dept. of CS \& IS, Goa Campus, Goa, India} \\ {\footnotesize Email: \{ssahay, h20160014\}@goa.bits-pilani.ac.in}}

\date{}

\maketitle

\begin{abstract}
	Due to continuous increase in the number of malware (according to AV-Test institute total $\sim 8 \times 10^8$ malware are already known, and every day they register $\sim 2.5 \times 10^4$ malware) and files in the 
computational devices, it is very important to design a system which not only effectively but can also efficiently
 detect the new or previously unseen malware to prevent/minimize the damages. Therefore, this paper  presents a novel group-wise approach for the efficient detection of  malware by parallelizing 
the classification using the power of GPGPU and shown that by using the Naive Bayes classifier the detection speed-up can be boosted up to 200x. The investigation also shows that the classification time increases significantly  with the number of features.
\vspace*{0.1cm}
~
\end{abstract}

\section{Introduction}
 The ubiquity of the Internet has engendered the prevalence of information
sharing among networked users and organizations, and in todays information era, most of the computing devices are connected to the Internet, which has rendered possible countless invasions of privacy/security worldwide from the malware ({\bf mal}icious soft{\bf ware}). In 1970 the first virus was created \cite{Bilar:2007:OPM:1359308.1359312}, and since then malware are not only continuously evolving with high
complexity to evade the available detection techniques, but also the new variants of malware
are increasing exponentially, as a consequence, malware defense is becoming a difficult task to protect the computational devices from it. The
use of malware for espionage, sophisticated cyber attacks, and other crimes motivated to develop an advanced method to combat
the threats/attacks from it \cite{Huda_2018} \cite{patent} \cite{DBLP:journals/corr/abs-1802-10135}\cite{Sharma2016AnEA}. However, due to the exponential increase in the number of malware (according to AV-Test institute total $\sim 8 \times 10^8$ malware are already  known, and every day they register $\sim 2.5 \times 10^4$ malware \cite{AV-test})), anti-malware industries not only facing major challenges to check the potential malicious content but to detect the malware efficiently. The reason behind these high volumes of malware is basically in the advancement of second-generation malware which can create millions of its variants by using different obfuscation techniques \cite{ashu1}.  The malware attack/threat are not only limited to individual boundaries, but they are highly skilled state-funded hackers writing customized malicious payloads to disrupt political, industrial working and military espionage \cite{symc17} \cite{stone} \cite{sans1}. The most high-profile, subversive incident was a series of intrusions against the Democratic Party in the US presidential election \cite{symc17}. 

\par In 2017 McAfee has more than 780 million malware samples in their database, and in the 3rd quarter of 2017 there was a 10\% increase in the number of the new malware,  in addition in the same quarter they have observed a 60\% increase in new mobile malware in the Android devices which are mainly due to increase in Android screen locking ransomware \cite{TRMacfee2017}. The Symantec 2017 Internet Security Threat report indicates that there were 357 million new malware variants \cite{symc17}. The recent Internet Security Threat Report from Symantec shows an increase of 88\% in overall malware variants \cite{symc18}. 
Hence, if adequate advancement in anti-malware technique is not achieved, consequences at this scale at which new malware are being developed can create fatal effects, and the results will be more severe then past. In this recently, various machine learning techniques have been proposed by authors \cite{allix} \cite{canto} \cite{ashu2} \cite{daniele}, which can enhance the capabilities of traditional malware detection system viz. signature matching technique, but with the use of a complex machine learning the detection time increases. Therefore, understanding the exponential increase in the number of malware released every year and files in the 
computational device, it is very important to design a system which not only effectively but can also efficiently detect the new or previously unseen malware to prevent/minimize the damages. Hence in this paper, we  present a novel efficient group-wise static malware analysis approach for the efficient detection of  malware by parallelizing 
the classification using the power of general-purpose graphics processing unit (GPGPU) and shown that by using the Naive Bayes classifier the detection speed-up can be boosted up to 200x. Accordingly, section 2 briefly discusses the related work done in this field. Section 3 describes the data preprocessing and how features are selected for the classification. Section 4  contains the experimental and the result analysis of our approach. Finally, we summarize our conclusion in Section 5.

\section{Related Work}

With the evolution of complex second-generation malware which can generate millions of its variant, 
the detection techniques have also been made significant  progress from the early day traditional
 signature matching to deep learning  techniques to improve the detection accuracy \cite{DBLP:journals/corr/abs-1801-02950}. In this 
recently Ashu et al. showed that group-wise classification 
of Windows malware in the range of 5 KB, the detection accuracy can be achieved up to 97.95\% \cite{Sharma2016AnEA}. 
Similarly, they have also shown that on an average 97.15\% detection
of Android malware can be achieved by  permission-based group-wise detection system \cite{10.1007/978-981-10-5508-9_20}. However, understanding the exponential growth of malware and the number of file in our system it is equally
important to focus on the design of an efficient malware detection system. In this 
Ciprian Pungila and Viorel Negru in 2012
 has proposed an efficient memory compression model for virus signature matching using GPGPU  \cite{pungila}. They were able to achieve 22 less memory utilization and 38 times higher bandwidth compared to their single-core implementation. In 2014, Che-Lun Hung et al., proposed a GPU based botnet detection technique \cite{Hung2014ParallelBD}. They implemented the network traffic reduction on GPU and were able to achieve eight times performance over CPU based traffic reduction. For Android devices, Manel Abdellatif et al.  in 2015 has designed and implemented a host-based parallel anti-malware based on mobile GPU \cite{Manel}. Their implementation was three times faster than the serial implementation on CPU. In 2016, to accelerate the statistical detection of zero-day malware Igor Korkin et al. has proposed a technique using CUDA-enabled GPU Hardware \cite{DBLP:journals/corr/KorkinN16}. In their work, they used GPU mainly for achieving speedup in memory forensic task. Recently, Radu Velea et al. has proposed a CPU/GPU based hybrid approach to accelerate pattern matching of the malware \cite{Velea2017CPUGPUHD}. In their work they found that the hybrid approach takes half of the time compared to the CPU implementation only and consumes 25\% less power.

\section{Data Preprocessing and Feature Selection}

For the experimental analysis, we downloaded 11,355 malware from the malicia-project dataset (one of author possess the dataset \cite{ashu2})) and collected 2967 benign programs (also verified from virustotal.com) from different windows operating systems. It has been observed that 97.18\% malware in the Malicia dataset is below 500 KB
\cite{Sharma2016AnEA}. Therefore, we took both the samples (i.e., malware and benign programs) which are below 500 KB, and left with 11,305 malware and 2360 benign executables for the analysis. Also, the investigation  by Ashu et al. shows that the variation in the size of the malware generated by malware kits viz. NGVCK, PS-MPC, and G2 do not vary by more than 5 KB range. Therefore, for efficient and effective classification we partitioned the datasets in 100 group each of 5 KB size. 

\par We selected the opcodes of the executable as a feature for the classification of malware because the difference in the opcode occurrence between the malware and benign executable differ in large \cite{Sahay:2016:GED:3066572.3066745}\cite{Sharma2016AnEA}. Therefore the prominent features i.e., opcodes from the data set which can differentiate the malware from benign programs are obtained as given by the Ashu et al. \cite{ashu2}\cite{Sharma2016AnEA} i.e.,  by normalizing the opcodes occurrence difference between the malware and benign executables for all the formed groups independently, and finally top k-features (opcodes) are selected from each group separately for the efficient and effective detection of the malware.

\section{Our Approach and Experimental Analysis}
\begin{figure}
	\centering
	\includegraphics[scale=.26]{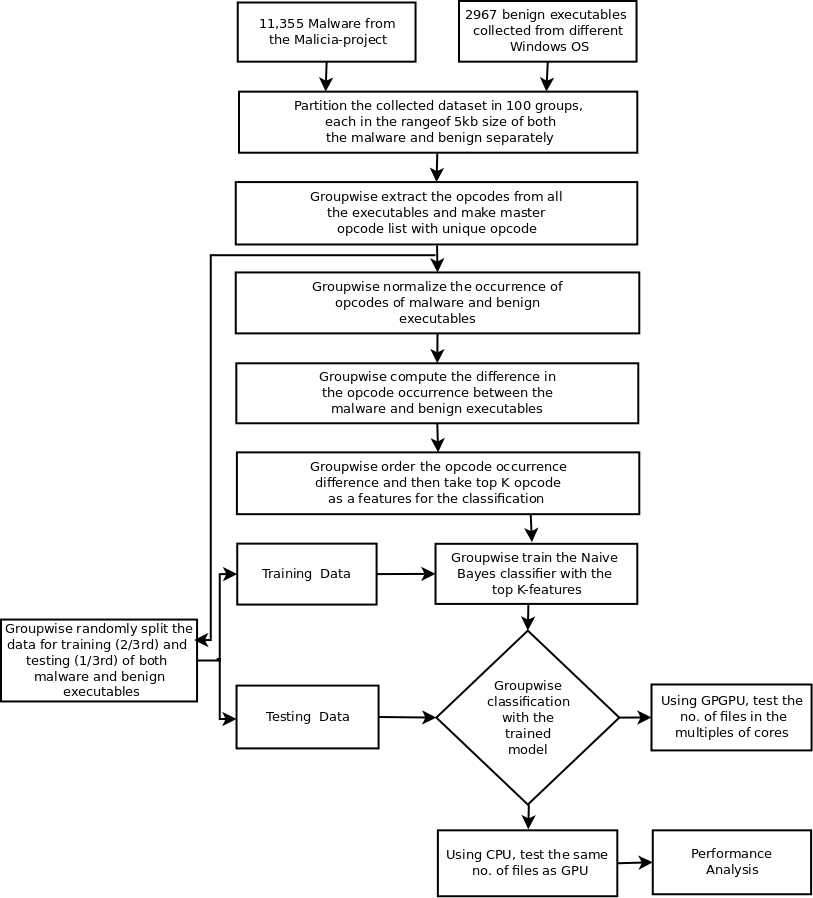}
	\caption{A schematic of our approach for the efficient detection of malware.}
\label{figure:fc}
\end{figure}

A schematic of the experimental analysis of our approach is shown in the fig. \ref{figure:fc}. For the purpose, we randomly split the dataset (containing only opcode occurrence of every executable) for the training and testing of the malicious and benign dataset

 \noindent separately (a conservative side as per the suggested norms to ensure optimal performance \cite{Guyon-1997}) in the ratio of 2:1 and used Naive Bayes classifier (as the paper focusses on the efficient detection of malware, not on the accuracy, therefore for simplicity we selected the Naive Bayes classifier) which assume strong class independence between different attributes under consideration, i.e., if the given set of  (opcodes in our case), A = $a_1, a_2, a_3,..., a_n$, then the Naive Bayes model computes posterior probability for target class C (malware/benign) and can be represented as \cite{ashu2}
$$P(C| a_1, a_2, a_3,..., a_n) = \frac{P(a_1, a_2, a_3,..., a_n |C)}{P(a_1, a_2, a_3,..., a_n)}$$
where, $P(C| a_1, a_2, a_3,..., a_n)$ is the posterior probability of an executable sample of belonging to class C. Hence one can calculate the posterior probabilities for the test executable, and if the malware class probability is higher then it is classified as malware otherwise it is labeled as benign.

The experiment has been conducted in Intel i7-7700HQ quad-core processor with a base frequency of 2.8 GHz, 8 GB RAM, Pascal architecture (GP107) based Nvidia 1050Ti GPU with 768 CUDA cores distributed across 6 SMP and 4GB GPU DRAM operating on a base speed of 1291 MHz. 

\par To improve the detection  efficiency, we trained the model  for all the groups independently with top k-features obtained from each group, except the group which has less than six files either in malware or benign. We find that 
5, 8, 61, 65, 97, 98 and 100th group have less then six files in either category (benign or malware). For the actual implementation if any 
group have less than the minimum set number of file, than that group file can be classified/tested with the next group trained model.

\par
\begin{figure}[H]
	\centering
	\includegraphics[scale=.49]{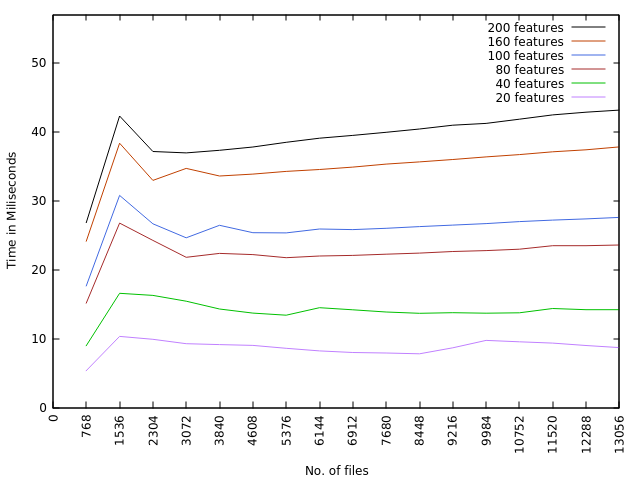}
	\caption{Time taken by the CPU to classify the files in the multiple of 768 files by varying the number features.}
\label{figure:cpu}
\end{figure}

\begin{figure}
	\centering
\includegraphics[scale=.49]{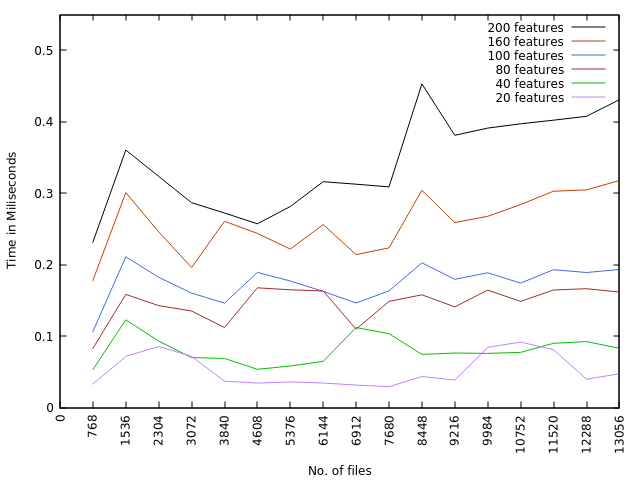}
	\caption{Time taken by the GPU to classify the files in the mulitple of 768 files by varying the number features. }
\label{figure:gpu}
	\end{figure}

 First we investigated the classification time taken by the CPU by selecting the top 20, 40, 80, 100, 160 and 200 features (figs. \ref{figure:cpu} and \ref{figure:gpu}) from each group and the

\noindent number of file in multiple of 768 (i.e., number of cores in the GPU), and the results obtained are shown in fig. \ref{figure:cpu}. Next, we find the time taken by GPU after distributing our trained model among all the 768 cores such that trained model of the particular group, the corresponding test file, and top k-features shall be
in same core so that parallelization of the tasks in GPU  can be optimally used. Then we observed the time taken by the GPU by giving the file in the multiple of 768 for the testing/classification and the results obtained are shown in fig. \ref{figure:gpu}.

\par
\begin{figure}
	\centering
\includegraphics[scale=.49]{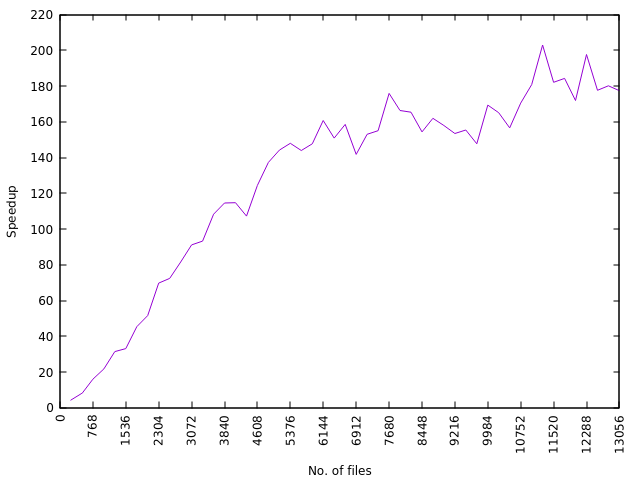}
	\caption{Classification speed-up due to parallelization of the task using the GPU.}
\label{figure:su}
	\end{figure}

 The analysis shows that the classification/testing time is also dependent on the number of features, and with the increase in the number of files in the multiple of number cores of GPU, the CPU proportionally take more time than GPU (figs. \ref{figure:cpu} and \ref{figure:gpu}). Therefore, experimented with various sets of features and found that the detection accuracy improves by increasing the number of features till top 200 features, after that there is no significant change in accuracy, and remains around $\sim$87\%. Therefore, we  investigated the speed-up with top 200 features (speed-up can be written as, Sp = Tc/Tp, where Tc is the time taken to execute the sequential program and Tp is the time taken to execute the program in parallel, i.e., in GPU with P number of cores \cite{parallel_computing}) and almost all the dataset (as our focus is on the efficiency not on the effectiveness of the classification) for the improvement
in the performance due the parallelization of the task using GPU that can be achieved from the given system, and the obtained result is shown in fig. \ref{figure:su}. We observed that the parallel implementation for the classification of malware using GPU by Naive Bayes algorithm is able to achieve speed-up up to 200 (not taking in account of the overhead involved in the processes).

\section{Conclusion}
We present a novel group-wise approach (to the best of our knowledge, this is the first paper that group-wise classifies the malware using the power of GPGPU) for the efficient detection of new or previously unseen malware by parallelizing the classification using the power of GPGPU and shown that by using the Naive Bayes classifier the classification speed-up can be boosted up to 200x (not taking in account of the overhead involved in the processes).  However, one has to study the performance using the classical  Random Forest classifier and Deep Learning methods for the efficient and high accuracy classification. Also, the trade-off between the efficiency and accuracy has to be investigated in-depth, i.e., optimal features selection (as classification time significantly increases with the number of features) to train the model after appropriately grouping the input data  for the detection of malware, and in this direction work is in progress.

\bibliographystyle{plain}
\bibliography{references}

\end{document}